\documentclass[twocolumn,prl,aps,showpacs,flushbottom]{revtex4-1}
\usepackage{xspace,amsmath,amsfonts,amsthm,amssymb,amsbsy,graphicx,color}

\begin{document}

\newtheorem{theo}{Theorem} \newtheorem{lemma}{Lemma}

\title{Real-time feedback control of a mesoscopic superposition}

\author{Kurt Jacobs,$^{1,2}$ Justin Finn,$^{1}$ and Sai Vinjanampathy$^{1}$}
\affiliation{$^{1}$Department of Physics, University of Massachusetts at Boston, 100 Morrissey
Blvd, Boston, MA 02125, USA \\
$^{2}$Hearne Institute for Theoretical Physics, Louisiana State
University, Baton Rouge, LA 70803, USA}

\begin{abstract}
We show that continuous real-time feedback can be used to track, control, and protect a mesoscopic superposition of two spatially separated wave-packets. The feedback protocol is enabled by an approximate state-estimator, and requires two continuous measurements, performed simultaneously. For nanomechanical and superconducting resonators, both measurements can be implemented by coupling the resonators to superconducting qubits.   
\end{abstract}

\pacs{85.85.+j, 42.50.Dv, 85.25.-j, 42.50.Lc} 
\maketitle

Methods for continuous-time tracking and control of continuous-variable quantum systems have to-date been restricted to localization about single points in energy or phase space, a direct quantum analogue of classical feedback control~\cite{DJ, Steck04, James04, Wiseman05, Geremia06}. One of the most striking features of quantum mechanics is the ability for systems to exist in spatially separated superpositions, and an important question is whether there is a potentially feasible way to use feedback to track and control such highly non-classical states, while preserving and protecting the coherence. The method for monitoring the square of position of a nano-mechanical resonator presented in~\cite{Jacobs09c}, along with the observation in~\cite{Jacobs09} that this measurement leaves evolving mesoscopic superpositions intact, have provided one of the tools necessary for realizing this task. 

For feedback to be able to preserve a superposition, it is not enough for the measurement to merely avoid destroying the coherence, it must also replace information that environmental noise removes. Although we will show that measurements of $x^2$ are not sufficient to do this, we can complement them with a quasi-continuous parity measurement, using a readily available off-resonant interaction, and this provides the missing information. The central element of this parity measurement was introduced in~\cite{Zippilli03} in a cavity-QED setting: in our solid-state setting it is not only more practical, but we are able to implement it in a quasi-continuous, rather than a discrete manner. The second element essential to  feedback control is an approximate state-estimator that can track the superposition in real-time. Using the estimator we develop below, we show that the state can be tracked, and the separation of the superposition controlled using the feedback cooling method introduced in~\cite{Steck04, Steck06}. 

We note that a previous method has been devised for protecting coherence in optical cavities. In a sequence of papers the group of Tombesi refined a feedback procedure in which a control system is correlated with the parity of the cavity field, and if a photon has been lost it is replaced by an adiabatic transfer from a single atom~\cite{Vitali97, Vitali98, Fortunato99, Zippilli03} (see also~\cite{Horoshko97}). This procedure is a very elegant example of coherent feedback control~\cite{Lloyd00, Nurdin09, Kerckhoff10}. While their protocol could be a useful tool in mesoscopic systems, it is quite distinct from our protocol here; it does not provide the observer with detailed phase-space information (and is thus restricted to preserving only the parity), or provide an efficient method to track the state.  We note also that our protocol does not employ a photon replacement operation. This is partly because our method separates the task of preserving coherence from that of manipulating the parity of the superposition (symmetric or anti-symmetric): the protocol performs the former, but the not the latter. The protection of coherence is our goal here since this is the key quantum feature of the state. 

%We have two primary purposes here. The first is to show that continuous-time state-estimation and feedback can track and preserve coherence between two points in phase space, thus bringing continuos-variable feedback out of the ``semiclassical'' realm and into this truly quantum domain. The second is to present a potentially feasible to perform such control with nano-electromechanical systems. 

The system that we wish to control is a single quantum oscillator (with frequency $\omega$, position $x$, and momentum $p$) in a superposition of two wave-packets that are well-separated in phase space. We will focus on nano-mechanical resonators, since preparing these in mesoscopic superpositions is a present goal of experimental work~\cite{Schwab05, Armour08a, Armour08b, LaHaye09, OConnell10}. It was shown in~\cite{Jacobs09} that a mesoscopic superposition will remain preserved by a measurement of $x^2$, so long as the position-space probability density for the superposition is symmetric in $x$. Further, a continuous measurement of $x^2$ will create such symmetric superposition states from an initial thermal mixture. This latter fact implies that an $x^2$ measurement must be able to extract parity information (at least for small phonon-number) in order to purify the initial mixture into a well-defined superposition.

We begin by deriving an approximate state-estimator for the $x^2$ measurement, which will also reveal how and when this measurement will extract parity information. This is achieved by constructing an ansatz for the state of the resonator, general enough to provide a good approximation throughout the evolution. We choose this ansatz to be a mixture of symmetric and antisymmetric superpositions of pure Gaussian wave-packets. If we denote by $|G_\pm \rangle$ the pair of general pure Gaussian states with centroids $(\pm \bar{x}, \pm \bar{p})$, variances $V_p$ and $V_x$, and symmetrized covariance $C$, then the superpositions  
\begin{equation}
   | \pm \rangle = \left( |G_+\rangle \pm |G_-\rangle \right) / \sqrt{2(1\pm \chi)} 
\end{equation}
are respectively a symmetric state (containing only an even number of photons) and an anti-symmetric state (containing an odd number). Here $\chi = \mbox{Re}[ \langle G_+ | G_- \rangle]$ is the overlap between the two Gaussians. Our full ansatz for the state of the resonator is then 
\begin{equation}
   \rho =  P | + \rangle \langle + | + (1-P) | - \rangle \langle - | ,  
\end{equation}
and contains the six parameters $\bar{x}, \bar{p}, V_x, V_p, C$ and $P$. This ansatz is based on the fact that the source of decoherence (the thermal bath) can be described in terms of phonon emissions and absorptions~\cite{Vitali97}. These change the state by one photon (toggle the symmetry), and correspond to operations linear in $x$ and $p$, so individually they preserve the ansatz. But since the observer does not have access to the times at which these random phonon events occur, he or she must average over all possible sequences of them. This averaging procedure no longer preserves the ansatz, because different sequences shift the wave-packets by different amounts, resulting in more terms in the density matrix. Nevertheless, so long as the measurement is strong enough compared to the rate of decoherence (in any case required for effective control), the even and odd components will remain nearly pure, and so the ansatz should provide a good approximation in this regime of ``good control''. 

The stochastic master equation (SME) that describes the continuous measurement of $x^2$ is~\cite{JacobsSteck06}  
\begin{equation}
   d\rho =  - k [M,[M,\rho]] dt + \sqrt{2k} ( M\rho + \rho M - 2\langle M\rangle \rho) dV ,  
   \label{sme}
\end{equation}
with $M = x^2$. The stochastic increment $dV$ is obtained by the observer from the continuous stream of measurement results, $r(t) = 4 k \langle M \rangle dt + \sqrt{2k} dW$, via the relation  
\begin{equation}
  dV = dr/\sqrt{2k} - \sqrt{8k} \langle M \rangle_{\mbox{\scriptsize e}} dt . 
\end{equation}
If the observer is solving the exact master equation, then her estimated value of $\langle M \rangle$ is simply $\langle M \rangle_{\mbox{\scriptsize e}} = \mbox{Tr}[\rho M]$, and $dV = dW$. If the observer is solving an approximate estimator, then $\langle M \rangle_{\mbox{\scriptsize e}}$ is the corresponding approximate estimate. We are assuming here that our continuous measurements are quantum-limited, and this requires quantum-limited amplifiers. While such amplifiers are not yet available for mesoscopic systems, recent progress in this area has been rapid, and it appears likely that such amplifiers will be realized in the near future~\cite{Bergeal10}. 

To derive the approximate state-estimator, an equation of motion for the six parameters, we substitute the ansatz into the SME, Eq.(\ref{sme}), and assume that the evolution preserves the Gaussian form of  $|G_\pm\rangle$. The resulting approximate state-estimator is 
\begin{eqnarray}
 d \bar{x}  & = &  (\bar{p} /m ) dt  + \xi \bar{x} V_x dV \\
 d \bar{p} & = & - m \omega^2 \bar{x} dt  + \xi \bar{x} C dV \\ 
 dV_x & = &  \left[ 2C/m - \xi^2 \bar{x}^2 V_x^2 \right]  dt + \xi V_x^2 dV  \\ 
 dC & = &  \left[ V_p/m -  m \omega^2 V_x  - \xi^2 \bar{x}^2 V_x C  \right] dt + \xi V_x C dV \\ 
 dV_p & = & - 2 m\omega^2 C  dt  + \xi \left[C^2 - \hbar^2/4 \right] dV \nonumber  \\ & & + \xi^2 \left[ (\hbar^2/4)( V_x + 2\bar{x}^2) - C^2 \bar{x}^2  \right] dt \\
% dV_p & = & - 2 m\omega^2 C  dt  + \xi \left[ \frac{\hbar^2}{4} - C^2 \right] dV \nonumber  \\ & & + \xi^2 \left[ \frac{\hbar^2}{4}( V_x + \bar{x}^2) + \left( \frac{\hbar^2}{4} - C^2 \right) \bar{x}^2  \right] dt \\
 dP & = & \sqrt{8k} \left( \langle x^2 \rangle^{\mbox{\scriptsize e}}_+ - \langle x^2 \rangle^{\mbox{\scriptsize e}}_- \right) P(1-P) dV , 
\end{eqnarray} 
where we have defined $\xi = 2\sqrt{8k}$, and the quantities $\langle x^2 \rangle^{\mbox{\scriptsize e}}_\pm$ are the estimated expectation values of $x^2$ for the states $|\pm\rangle$. So long as the wave-packets are well-separated ($\chi=0$) we have $ \langle x^2 \rangle^{\mbox{\scriptsize e}}_+ = \langle x^2 \rangle^{\mbox{\scriptsize e}}_- = V_x + \bar{x}^2$. 

Note that it is the mixing probability, $P$, that determines the level of coherence: $P=1$ and $P=0$ describe perfect coherence, and $P=1/2$ denotes zero coherence. To preserve coherence the feedback protocol must keep the absolute value of the parity, $|\mathcal{P}| = |2P - 1|$, close to unity. Examining the equation for $P$, we see that this is exactly the equation for a classical continuous measurement that distinguishes between the two states of a single bit. The rate at which this information is extracted can be read off directly, and is $\sqrt{8k} \left( \langle x^2 \rangle^{\mbox{\scriptsize e}}_+ - \langle x^2 \rangle^{\mbox{\scriptsize e}}_- \right)$. This tells us that the measurement provides parity information so long as $\langle x^2 \rangle^{\mbox{\scriptsize e}}_+ \not= \langle x^2 \rangle^{\mbox{\scriptsize e}}_-$. This is true only if the Gaussian states in the superposition are close enough in phase space to significantly overlap. Our position measurement will effectively track the location of the wave-packets, but will not purify a mixture of the even and odd superpositions once they are well-separated in phase space. In fact we conclude that no practicable continuous symmetric measurement of position can determine the parity of these states --- while the packets cross each other (in position space) as they pass $x=0$ every half period, the spatial period of the oscillations in the overlap become increasingly rapid, so that  impractically fine resolution in position would be required to distinguish odd from even.  

The thermal bath destroys parity information, reducing $|\mathcal{P}|$ and thus the coherence over time. The measurement must continually replace this information to preserve the coherence. To solve this problem we now show how to make a quasi-continuous parity measurement, which can be performed in conjunction with the $x^2$ measurement. To do this we couple the resonator off-resonantly to a single superconducting qubit~\cite{Irish03}. The interaction is given by $H_{\mbox{\scriptsize int}} = \hbar g \sigma_z a^\dagger a$, where $\sigma_z$ is the Pauli operator for the qubit. Let us denote the eigen-states of the $\sigma_x$ operator as $|+\rangle$ and $|-\rangle$. If we start the qubit in state $|+\rangle$, and switch on the interaction for a time $\tau = \pi/g$, then the qubit is flipped to state $|-\rangle$ if and only if the resonator has an odd number of photons. This correlates the qubit perfectly with the resonator parity. If we now make a continuous measurement of $\sigma_x$ with strength $\mu$ --- which is described by Eq.(\ref{sme}) with $M = \sigma_x$ and $k = \mu$ --- then it translates to a measurement of the resonator parity while the correlation is maintained. The thermal damping of the resonator slowly de-correlates the two systems, and we must take this into account when determining the evolution of $P$. If we denote the probability that the qubit is in state $|+\rangle$ by $P_+$, then the continuous measurement of the qubit reduces to  
\begin{equation} 
   dP_+ = \sqrt{2\mu}P_+(1-P_+) dU ,  
   \label{qmeas} 
\end{equation} 
where $dU$ is obtained from the  measurement record, $r_x(t)$, by $dU = dr_x(t)/\sqrt{2\mu} - \sqrt{8\mu}\langle \sigma_x \rangle dt$.  In our case, the degree of correlation between the qubit and the parity is fully determined by the conditional probability for the resonator to have positive parity, given that the qubit is in state $|+\rangle$. Denoting this conditional probability by $1 - \beta$, the observer's estimated mixing probability is $P = (1-\beta) P_+  + \beta (1-P_+)$. Since the thermal damping de-correlates the qubit, we reset and re-correlate the qubit at regular intervals. We obtain good tracking of the parity if we allow the estimator to assume that during the correlation step, $P$ decays due to the thermal damping, but at the end of the step, $\beta=0$ (perfect correlation), so that the estimator can reset $P_+$ to the current value of $P$. Between re-correlations, the evolution of $\beta$ is determined by the thermal damping, and we give this below. We do not take into account relaxation (or dephasing) of the qubit in the above analysis, because simple quantum (or in fact classical) error correction can be used on the qubit to eliminate the effects of relaxation on the measurement. 

Both the $x^2$ and $\sigma_z$ measurements must be switched-off while the qubit is correlated with the resonator parity, since they will interfere with it. We thus make the correlation step short compared to the effective frequency of the resonator, which is feasible since the effective frequency can be relatively low (e.g.\ $250~\mbox{kHz}$, see below). 

Before we simulate our state-estimation procedure, we need to know how the thermal damping of the resonator changes our six variables, so as to include this dynamics in the state-estimator. These equations of motion are necessarily an approximation, because as mentioned above the damping does not preserve the ansatz. In the well-separated regime ($\chi=0$), our approximate damping  equations are $\dot{\bar{x}} = -(\gamma/2) \bar{x}$, $\dot{\bar{p}} = -(\gamma/2) \bar{p}$, $\dot{V}_x =  - \gamma (V_x - V_x^{\mbox{\scriptsize Th}}) $, $\dot{V}_p =  - \gamma ( V_p - V_p^{\mbox{\scriptsize Th}} )$, and $\dot{C} = - 2\gamma C$. In these equations $\gamma$ is the damping rate of the resonator, and $V_x^{\mbox{\scriptsize Th}} \equiv (2 n_T + 1) \Delta_x^2$ and $V_p^{\mbox{\scriptsize Th}} \equiv (2 n_T + 1) \Delta_p^2$  are the variances of the harmonic oscillator at temperature $T$. The quantities $\Delta_x^2 = \hbar/(2m\omega)$ and $\Delta_p^2 = \hbar m \omega/2$, are, respectively, the variances of $x$ and $p$ in the ground state.  The parameter $n_T$ is the average number of phonons that the resonator would have if it were at the bath temperature $T$. The evolution of the qubit/resonator correlation between correlation steps is 
\begin{equation}
   \dot{\beta} = - \gamma \left[ \tilde{V_x} + \tilde{\bar{x}} + \tilde{V_p} + \tilde{\bar{p}} - 1 \right] (\beta - 1/2) ,  
\end{equation}
where the tildes indicate that the means and variances are those of the dimensionless variables $\tilde{x} = (a+a^\dagger)/\sqrt{2}$ and $\tilde{p} = -i(a-a^\dagger)/\sqrt{2}$. Note that the expression in the square brackets is merely the estimated value of $2 \langle a^\dagger a \rangle$.  

\begin{figure}[t]
\leavevmode\includegraphics[width=1.0\hsize]{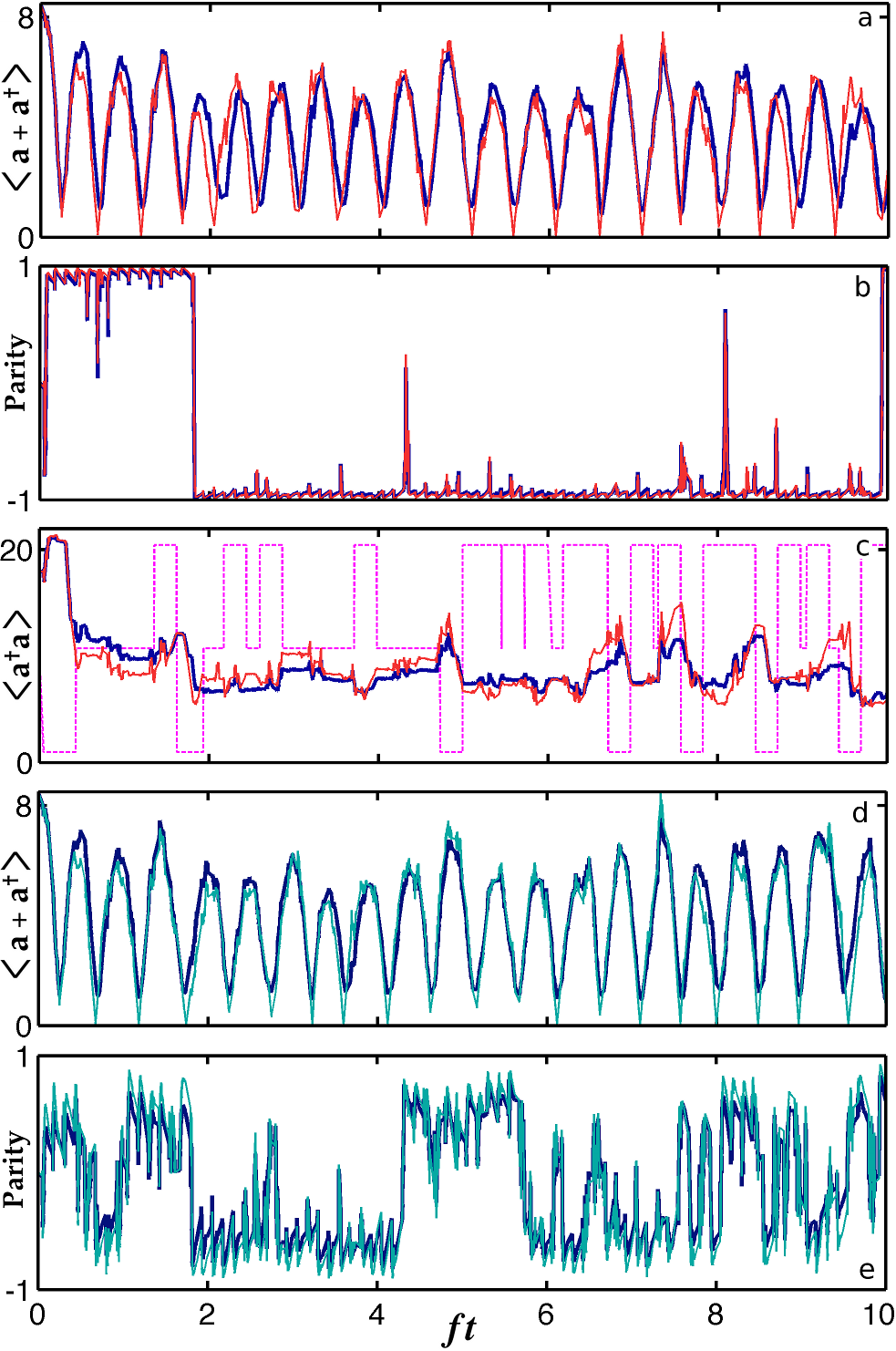}
\caption{(color online) Time series' for the true and estimated values of the evolution of various  observables of the resonator under the feedback control protocol. (a), (b) and (c) have resonator quality factor $Q=200$, and temperature $T=0$. (d) and (e) have $Q=500$, and $T = 20~\mbox{mK}$ ($n_T = 12$). The dark (blue) solid lines are the true values, and the lighter solid lines (red or blue) are the estimated values. In (c) the dashed line indicates when the feedback is heating (high value), dormant, or cooling (low value) the system.} 
\label{fig1}
\end{figure}

We now perform simulations to verify that our protocol is able to track and preserve the superposition, as well as control the phase-space separation of its two components. We do this using experimentally realistic parameters, as follows. Nano-resonators typically have frequencies $f = \omega/(2\pi)$ in the range $10 - 100~\mbox{Mhz}$. By modulating at frequency $\Delta$ the coupling of the resonator with the probe qubit mediating the $x^2$ measurement, we can reduce the oscillation frequency (from the point of view of the measurement and thus the observer) to $\nu = \omega - \Delta$. This means that the state-estimation need only work on the timescale of $\nu$, rather than $\omega$. We chose $\omega/(2\pi) = 50~\mbox{MHz}$, and $\nu/(2\pi) = 250~\mbox{kHz}$. Since nano-resonators typically have Q factors of $10^5$, the damping rate is $\gamma/(2\pi) = 5 \times 10^3~\mbox{s}^{-1}$, and this gives an effective Q  of $ \nu/\gamma = 500$. Both the $x^2$ measurement and the energy interaction $H_{\mbox{\scriptsize int}}$ are obtained by using an off-resonant (perturbative, 2$^{\mbox{\scriptsize \textit{nd}}}$-order) interaction with a qubit~\cite{Jacobs09}. The maximum rate of direct coupling between qubits and resonators is $\sim 10^8~\mbox{s}^{-1}$~\cite{Armour02}, so that of 2$^{\mbox{\scriptsize \textit{nd}}}$-order interactions is $\sim 10^7~\mbox{s}^{-1}$. To fix the correlation step at $1/20$ of the effective resonator period, we need the interaction rate $g = 12.5\nu \approx 8\times 10^6~\mbox{s}^{-1}$. The rates of the $x^2$ and $\sigma_z$ measurements should ideally be much larger than $\gamma n_T$, and the former should be less than or $\nu$ to avoid undue heating. For the following simulations we chose $k = \nu/(200\pi)$ and two values for $\mu$, $\mu = \nu$ and $\mu = \nu/2$. 

To simulate the continuous measurements of $x^2$ and $\sigma_z$ in the presence of thermal damping we use the quantum Monte-Carlo method described in~\cite{Jacobs10}. We also use the approximate state-estimator to implement the feedback cooling method devised in~\cite{Steck04, Steck06} to stabilize the amplitude (separation) of the superposition. The cooling method involves raising the potential (increasing the frequency of the oscillator) when the estimated value of $\langle x^2 \rangle$ is at a minimum, and reducing the frequency when it is at a maximum. To determine when $\langle x^2 \rangle$ has reached an extremal point, we fit a quadratic to a smoothed version of the time-series for the estimated $\langle x^2 \rangle$, looking back for one twentieth of a period. In our simulations we stabilize the state by applying cooling (heating) --- the latter using the reverse algorithm --- when the estimated average phonon number, $\langle n \rangle_{\mbox{\scriptsize e}}$, is greater than $25$ (less than $16$). This feedback stabilization is turned off when $16 < \langle n \rangle_{\mbox{\scriptsize e}} < 20$. 

In Fig.~\ref{fig1}(a)-(c) we simulate the estimation and control process using the parameters above, but with a lower quality resonator ($Q = 200$), and with the temperature set to zero. The initial state is an equal mixture of the even and odd superposition states, with $(\bar{x},\bar{p},V_x,V_p,C) = (6\sqrt{2}\Delta_x,0,\Delta_x^2,\Delta_p^2,0)$. Here we start the estimator with the correct values of the parameters, but we have verified that if the initial guess is wrong, the estimator locks on in two or three oscillation periods. The estimator appears to be quite robust, but to increase robustness our estimator truncates the measurement noise if it is outside 4 standard deviations, and resets the variances if one of the following conditions occurs: $V_x < 0,  V_p < 0, V_xV_p < 0.9(C^2 + 0.25\hbar^2), V_x > 12 \Delta_x^2$, or $V_p > 12 \Delta_p^2$. In Figs.~\ref{fig1}(a), (b), and (c) we show the true and estimated values of, respectively, the absolute value of the scaled position, parity, and average phonon number as functions of time, for a single run. These plots show that the estimator tracks the location and parity of the state well, and shows the degree to which the energy is stabilized by the feedback. The parity time-series also shows that the measurement is effective at keeping the state highly pure, and exhibits the quantum jumps in parity. These jumps are due to the interplay of the strong quasi-continuous measurement and the much weaker thermal decoherence (phonon damping) --- the measurement is able to reveal the phonon emission events that flip the parity.  

In Fig.~\ref{fig1}(d) and (e) we simulate the process again, but this time with $Q = 500$, and set the temperature to be $20~\mbox{mK}$. For our $50~\mbox{MHz}$ resonator this means that the thermal occupation number is $n_T = 12$. We see that the estimator still tracks well the position and the parity. But this time, due to the increase in $n_T$, the measurement is not able to maintain the purity as effectively as before. Nevertheless, the estimator still reveals the quantum jumps in the parity due to the thermal environment. 

{\em Acknowledgments:} KJ thanks Klaus M\/{o}lmer for helpful discussions. This work was supported by the NSF under Project Nos. PHY-0902906 and PHY-1005571, the Hearne Institute, the Army Research Office and the Intelligence Advanced Research Projects Activity. 

%\bibliography{report}

%Merlin.mbs v4.21 2009-07-09.
%

\end{document}